\documentclass[aps,prd,nofootinbib,floatfix,twocolumn,groupedaddress]{revtex4-1}
\usepackage{setspace}
\usepackage{graphicx}
\usepackage{dcolumn}
\usepackage{multirow}
\usepackage{stackrel}
\usepackage{subfigure}
\usepackage{times,mathptm}
\usepackage{float}
\usepackage{color}
\usepackage{amsmath,amsfonts}
\usepackage{mathptmx}
\usepackage{mathrsfs}
\usepackage{bbm}
\usepackage{bm}
\usepackage{xfrac}

\newcommand{\beq}{\begin{equation}}
\newcommand{\eeq}{\end{equation}} 
\newcommand{\bea}{\begin{eqnarray}}
\newcommand{\eea}{\end{eqnarray}} 
\newcommand{\qbar}{\overline{q}}

\def\lsim{\mathrel{\rlap{\lower4pt\hbox{\hskip1pt$\sim$}}
    \raise1pt\hbox{$<$}}}
\def\gsim{\mathrel{\rlap{\lower4pt\hbox{\hskip1pt$\sim$}}
    \raise1pt\hbox{$>$}}}

\renewcommand{\d}{\delta}

\renewcommand{\l}{\lambda}

\renewcommand{\b}{\beta}

\newcommand{\tr}{\text{Tr}}

\newcommand{\vx}{{\vec{x}}}
\newcommand{\vy}{{\vec{y}}}
\newcommand{\vz}{\vec{z}}

\newcommand{\vk}{{\vec{k}}}

\renewcommand{\vr}{\vec{r}}
\newcommand{\vR}{\vec{R}}
\newcommand{\ve}{\hat{\vec{e}}}

\newcommand{\m}{\mu}
\newcommand{\q}{\overline{q}}
\newcommand{\g}{\gamma}
\renewcommand{\r}{\rho}

\newcommand{\s}{\sigma}

\newcommand{\vp}{\vec{p}}

\newcommand{\N}{{\cal N}}

\newcommand{\oh}{\frac{1}{2}}
\newcommand{\oq}{\frac{1}{4}}

\newcommand{\dg}{\dagger}
\newcommand{\non}{\nonumber}

\newcommand{\rf}[1]{(\ref{#1})}
\newcommand{\ra}{\rightarrow}
\newcommand{\pa}{\partial}
\renewcommand{\vec}[1]{\bm #1}

\usepackage{ulem}

\begin{document}

\title{The Coulomb flux tube on the lattice} 

\bigskip
\bigskip

\author{Kristian Chung}
\author{Jeff Greensite}

\affiliation{\singlespacing Physics and Astronomy Department, San Francisco State
University,   \\ San Francisco, CA~94132, USA}
%

\begin{abstract}

\singlespacing

        In Coulomb gauge a longitudinal electric field is generated instantaneously with the creation of a static quark-antiquark pair.  The field due to the quarks is a sum of two contributions, one from the quark and one from the antiquark, and there is no obvious reason that this sum should fall off exponentially with distance from the sources.  We show here, however, from numerical simulations in pure SU(2) lattice gauge theory, that the color Coulomb electric field does in fact fall off exponentially with transverse distance away from a line joining static quark-antiquark sources, indicating the existence of a color Coulomb flux tube, and the absence of long-range Coulomb dipole fields.  

\end{abstract}

\maketitle

\singlespacing

\section{\label{intro} Introduction}

    Coulomb gauge has been used in many studies of quark confinement, beginning with the seminal work of Gribov 
\cite{Gribov:1977wm}, later elaborated by Zwanziger \cite{Zwanziger:1998ez}.  In this gauge the color Coulomb potential (defined below) is confining, and there is some hope that this confining behaviour can be derived or understood analytically, e.g.\ by Schwinger-Dyson equations, variational methods, or some other approach.   A sample of work along these lines is found in 
\cite{Szczepaniak:2001rg,Szczepaniak:2003ve,Feuchter:2004mk,Epple:2006hv,Zwanziger:2003de,Alkofer:2009dm,Golterman:2012dx}.
Coulomb gauge also has the advantage that physical states are obtained by operating on the vacuum with local field operators.   This allows us to define what is meant by ``constituent'' gluons in hadronic states, and to construct e.g.\ glueball states by operating on the
vacuum with $A$-field operators in the appropriate combinations of spin and parity.

    The color Coulomb potential $V_C(R)$ is the interaction energy of the state $\Psi_{\qbar q}$ generated by quark-antiquark creation operators acting on the ground state, i.e.
\bea
             {\cal E}_C(R)  &=& \langle \Psi_{\qbar q}|H|\Psi_{\qbar q} \rangle \non \\
                          &=&  V_C(R) + {\cal E}_0 \ ,
\eea
where $H$ is the Coulomb gauge Hamiltonian and ${\cal E}_0$ is an $R$-independent constant.  We will consider static quarks in the
infinite mass limit, with the quark located at position $\vR_1$, the antiquark at position $\vR_2$ ,and the ${\qbar q}$ state is
\bea
           |\Psi_{\qbar q}\rangle &=&  \N \int {d^3k_1 \over (2\pi)^3}  {d^3k_2 \over (2\pi)^3} b^{\dg \s}(k_1,\l_1) d^{\dg \s}(k_2,\l_2)  \non \\             
      & & \qquad \qquad \times       e^{-i(\vk_1 \cdot \vR_1 + \vk_2 \cdot \vR_2)}  |\Psi_0 \rangle \ ,
\label{Psi}
\eea
where $\Psi_0$ is the ground state,  $b^\dg, d^\dg$ are quark and antiquark creation operators,  $\s$ is a color index, $\N$ is a normalization constant, and the polarizations $\l_{1,2}$ are unimportant in what follows. For convenience we take $\vR_1$ to be the origin, and $\vR_2 = \vR= R \ve_x$ to lie along the $x$-axis. It is well known from lattice simulations \cite{Greensite:2003xf,Greensite:2004ke,Nakagawa:2006fk,Greensite:2014bua,Burgio:2012bk,Voigt:2008rr} that $V_C(R)$ is a linearly confining potential. What has not  
been investigated up to now is the spatial distribution of the color Coulomb field which gives rise to this potential.  

  The reason that there is any $R$-dependence at all in the energy expectation value is due to the fact that, in Coulomb gauge, the creation of charged sources is always accompanied, because of the Gauss law constraint, with an associated longitudinal electric field.  To briefly
review this point: In Coulomb gauge the dynamical degrees of freedom are the transverse $A,E^{tr}$ fields.  Separating the color
electric field into a transverse and longitudinal part,  $E = E^{tr} + E_L$ where $E_L = -\nabla \phi$, the Gauss law constraint
$D_k E_k =\r_q$ becomes
\beq
            -\pa_i D_i \phi = \r_q + \r_g \ ,
\eeq
where $D_i$ is the covariant derivative, and
\beq
          \r^a_q = g\qbar T^a \g_0 q   ~~~,~~~ \r^a_g = g f^{abc} E_k^{tr,b} A_k^c
\eeq
are the color charge densities due to the quark and gauge fields, respectively, with $T^a$ the generators of the Lie algebra,
and $f^{abc}$ the structure constants.  Let
\beq 
            G^{ab}(\vx,\vy;A)  =   \left({1 \over - \pa_i D_i(A)}\right)^{ab}_{\vx \vy} 
\eeq
be the inverse of the Faddeev-Popov operator.  Then
the longitudinal electric field $E_L = -\nabla \phi$ is determined to be
\beq
          \vec{E}^a_L(\vx,A,\rho) =   -\vec{\nabla}_x \int d^3y ~  G^{ab}(\vx,\vy;A) (\r_q^b(y)+\r^b_g(y)) \ .
\label{Elong}
\eeq
We are interested in the part of $E_L$  which is generated by the quark-antiquark sources, namely
\beq
\vec{E}^a_{L,q\qbar}(\vx,A,\r_q) =   -\vec{\nabla}_x \int d^3y ~  G^{ab}(\vx,\vy;A) \r_q^b(y) \ .
\eeq
Now suppose $A_i$ is a typical vacuum fluctuation, where the word ``typical'' is best defined with a lattice regularization:  these
are thermalized configurations generated by the lattice Monte Carlo procedure, transformed to Coulomb gauge.   Squaring
$\vec{E}^a_{L,q\qbar}(\vx,A,\r_q)$, summing over the color index $a$, and taking the expectation value of the matter field 
color charge densities in the massive quark-antiquark state, a straightforward calculation leads to the matter contribution
\bea
        E^2_{L,q\qbar}(\vx,A)   &=&     {g^2 \over 2N_c} \Bigl(  \nabla_x G^{ab}(\vx,0;A) \cdot \nabla_x G^{ab}(\vx,0;A) \non \\
  & &  \quad +  \nabla_x G^{ab}(\vx,\vR;A) \cdot \nabla_x G^{ab}(\vx,\vR;A) \non \\                    
  & &  \quad        - 2 \nabla_x G^{ab}(\vx,0;A) \cdot \nabla_x G^{ab}(\vx,\vR;A)   \Bigr)      \ , 
\label{E2}
\eea
where $N_c$ is the number of colors.
It seems unlikely that $G^{ab}(\vx,\vy,A)$ would fall exponentially with $|\vx-\vy|$ for typical vacuum configurations.   
In that case it is hard to see how the color Coulomb potential, which depends on the interaction kernel
\beq
          K^{ab}(|\vx-\vy|)= \int d^3z \left\langle   \left( G^{ac}(\vx,\vz,A) (-\nabla^2)_{\vz}  G^{cb}(\vz,\vy,A) \right) \right\rangle \ ,
 \eeq
could rise linearly at large $|\vx-\vy|)$.  Moreover, the expectation value of the Fourier transform of $G^{ab}(\vx,\vy,A)$, which is the momentum-space ghost propagator $G^{ab}(\vk)$, has been computed in lattice Monte Carlo simulations, both in SU(2) 
\cite{Burgio:2008jr,Burgio:2012bk,Langfeld:2004qs} and SU(3) \cite{Nakagawa:2009zf}  pure gauge theory, with the result
\beq
              G^{ab}(\vk) = \langle G^{ab}(\vk,A) \rangle \sim {\d^{ab} \over |\vk|^{2.44}}
\label{ghost}
\eeq
in the infrared, corresponding to an asymptotic behavior $G^{ab}(r) \sim \d^{ab}/r^{0.56}$ in position space.  So it is reasonable to 
assume some power-law falloff of $G^{ab}(\vx,\vy,A)$ with separation  $|\vx-\vy|$,
for typical vacuum fluctuations $A$.  Then, unless there are very delicate cancellations among the terms in \rf{E2}, one would expect a power law falloff  for $E^2_L(\vx,A)$, as the distance of point $\vx$ from the $\qbar q$ sources increases.  This would imply a long-range color Coulomb dipole field in the physical state $\Psi_{\qbar q}$.

    It should be emphasized that $\Psi_{\qbar q}$ is not the minimal energy state containing a static quark-antiquark pair.  For that reason $V_C(R)$ is clearly an upper bound on the potential $V(R)$ of a static quark-antiquark pair, and if the static
quark-antiquark potential is confining, then so is the color Coulomb potential $V_C(R)$ (a point first made in \cite{Zwanziger:2002sh}).  In fact,  lattice
simulations  \cite{Nakagawa:2006fk,Greensite:2014bua} show that the color Coulomb potential in SU(3) pure gauge theory is about a factor of four greater than
the usual asymptotic string tension.  If one would begin with the physical state \rf{Psi} and let it evolve in Euclidean time, then the state will evolve to the minimal energy state with potential $V(R)$, and the initial color Coulomb electric field will evolve into the standard flux tube configuration.  It has been suggested \cite{Greensite:2015nea} that in Coulomb gauge the minimal energy flux tube state is best understood in the framework of the gluon chain model \cite{Greensite:2001nx}, where we consider more general states of the form
\bea
   |\Psi\rangle &=& \int \prod_{i=1}^n d^3x_i ~\Psi_{k_1\ldots k_n}(x_1,x_2,\ldots,x_n) \non \\
      & & \times \q^{+}(0) A_{k_1}(x_1) A_{k_2}(x_2)
         \ldots A_{k_n}(x_n) q^{+}(R) |\Psi_0 \rangle \ ,
\label{chain}
\eea
and the order of color indices of the $A$ fields is correlated with position in the chain.  In principle such states can reduce the Coulomb string tension to the asymptotic string tension; the details can be found in \cite{Greensite:2014bua,Greensite:2015nea}.  In this article, however, we will be concerned with the distribution of the Coulomb electric field associated with the state $\Psi_{\q q}$, and the question we address here is whether this color dipole gives rise to a long range Coulomb field, or whether instead the Coulomb electric field is somehow collimated from the moment of creation of the static quark-antiquark pair, even before that field has a chance to evolve into a minimal energy flux tube.

\section{\label{setup} Lattice Setup}

   We work in the framework of the Euclidean path integral of SU(2) lattice gauge theory in Coulomb gauge
\bea
          Z &=& \int DA_\m \d(\nabla \cdot A) M[A] e^{-S_{YM}}  \ ,
\eea
where $M[A]$ is the Faddeev-Popov determinant in Coulomb gauge, and in this article we will neglect the issue of Gribov copies.  To compute the Coulomb potential,
let
\beq
              L_t(\vx) \equiv T\exp\left[ig\int_0^t dt' A_4(\vx,t') \right] \ .
\eeq
Then the Coulomb energy is obtained from the logarithmic time derivative 
\cite{Greensite:2003xf,Greensite:2004ke}
\beq
  {\cal E}_C(R)  = - \lim_{t\ra 0} {d \over dt} \log  \big\langle \tr[L_t({\bf 0}) L_t^\dg({\bf R})] \big\rangle \ ,
\label{logtime}
\eeq     
while the minimal energy of static quark-antiquark state is obtained in the opposite limit
\beq
            {\cal E}_{min}(R) = - \lim_{t\ra \infty} {d \over dt} \log  \big\langle \tr[L_t({\bf 0}) L_t^\dg({\bf R})] \big\rangle \ .
\eeq
Now in Coulomb gauge the $\langle A_4 A_4 \rangle$ correlator has an instantaneous part
\bea
         D^{ab}_{44}(x-y) &=& \langle A_4^a(x) A_4^b(y) \rangle \non \\
                                    &=& \d^{ab} D(\vx-\vy) \d(x_0-y_0) + P^{ab}(x-y) \ ,
\eea
where $P^{ab}(x-y)$ is the non-instantaneous part.  It was shown by Zwanziger \cite{Zwanziger:1998ez,Cucchieri:2000hv} 
that both $g^2  D^{ab}_{44}(x-y)$ and $g^2 D(\vx-\vy)$ are renormalization group invariant.  Expanding $L_t$  in a power series and extracting the $R$-dependent part of ${\cal E}_C(R)$, it is clear that
\beq
          V_C(R) = g^2 C_F  D(R)        \ ,
\eeq 
where $C_F$ is the quadratic Casimir of the fundamental representation.

\begin{figure}[htb]
 \includegraphics[scale=0.6]{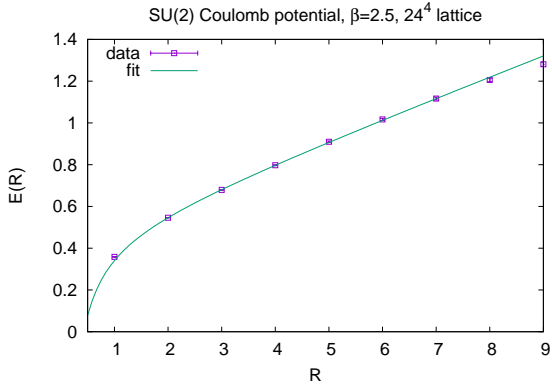}
 \caption{Coulomb interaction energy vs.\ on-axis quark separation for a static quark-antiquark pair in SU(2) pure gauge 
 theory at $\b=2.5$ on a $24^4$ lattice volume. All data is in lattice units.  The solid line is the best linear + $1/R$ fit to the data in the range $2\le R \le 8$, which in this case is ${\cal E}(R)= 0.094R - 0.27/R + 0.49$.  Error bars are smaller than than the symbol for the data
 points.}
 \label{prop}
 \end{figure}

   The lattice version of \rf{logtime} in SU($N$) pure gauge theory is the logarithm of the equal times timelike link correlator
\beq
            {\cal E}_C(R_L) = -    \log \big\langle {1\over N}\tr[U_0({\bf 0},0) U_0^\dg({\bf R}_L,0)] \big\rangle \ ,
\label{Llogtime}
\eeq
where $R_L$ is in lattice units, and in this form ${\cal E}_C(R_L)$ has been computed in numerical simulations.   Dividing by the lattice spacing to convert to physical units, it was found that  
\beq
       {\cal E}^{phys}_C(R) = \s_c(\b) R - {\g(\b) \over R} + {c(\b) \over a(\b)} \ ,
\eeq
where $a(\b)$ is the lattice spacing and $R=R_L a(\b)$.  In physical units the Coulomb string tension $\s_c(\b)$, and the dimensionless constants $\g(\b),c(\b)$, appear to have finite non-zero limits as ${\b \ra \infty}$, with the Coulomb string tension approximately four times larger than the asymptotic string tension, as shown in the SU(3) Coulomb gauge lattice simulations of ref.\ \cite{Greensite:2014bua}.  An intriguing fact is that $\g$ appears to go to $\pi/12$ in the continuum limit, which is the 
L\"uscher value expected for the QCD flux tube.  That could be a numerical coincidence, although this value of $\g$ is also roughly consistent with a best fit of our $SU(2)$ on-axis data for ${\cal E}_C(R_L)$ at $\b=2.5$, shown in Fig.\ \ref{prop}.  If this proximity to
$\pi/12$ is not a coincidence, and $\g$ does indeed have a string origin of some kind, that would be interesting to know.  This is part of our motivation to study the spatial distribution of the Coulomb electric field due to static quark-antiquark charges.
 
    For our purposes it is sufficient to study how the color electric field depends on the transverse distance away from the midpoint of a
line joining the quark and antiquark.   Let the quark and antiquark lie along the $x$-axis, say, with $\ve_x$ and $\ve_y$ unit vectors in
the $x,y$ directions, and define
\beq
            \vp = \oh R \ve_x + y \ve_y \ .
\eeq
The quantity we wish to compute is the contribution to the $x$-component $\langle \tr E_x^2 \rangle$ due to the quark-antiquark pair, i.e.
\bea
& & \langle \Psi_{\qbar q}| \tr E_x^2 (\vp) |\Psi_{\qbar q} \rangle - 
            \langle \Psi_0 | \tr E_x^2 |\Psi_0 \rangle 
\non \\
       & &  =   \lim_{t\ra 0} {\int DA \d(\nabla \cdot A) M[A] \tr[L_t({\bf 0}) L_t^\dg({\bf R})]  \{- \tr E_x^2(\vp, \oh t)\} e^{-S_{YM}}\over 
    \int DA \d(\nabla \cdot A) M[A]\tr[L_t({\bf 0}) L_t^\dg({\bf R})] )e^{-S_{YM}}}
\non \\
    & & \qquad - {1\over Z} \int DA   \d(\nabla \cdot A) M[A] \{- \tr E_x^2)\} e^{-S_{YM}}  \ .
\label{Qc}
\eea
\begin{figure}[htb]
\includegraphics[scale=1.0]{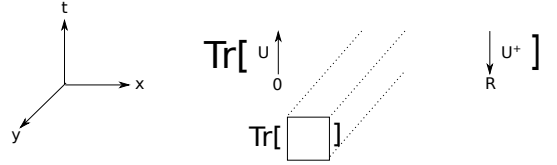}
\caption{The observable for the calculation of the $x$-component of the color electric energy density $Q(R,y)$, generated by a quark-antiquark pair along the $x$-axis separated by distance $R$, as a function of the transverse distance $y$ away from the midpoint. $U,U^+$ denote timelike link variables at equal times.}
\label{corr}
\end{figure}
The lattice version of \rf{Qc} is $4 Q(R,y)$, where
\bea
         Q(R,y) = { \langle \tr[U_0({\bf 0},0) U_0^\dg({\bf R}_L,0)]  \oh \tr U_P(\vp,0)\rangle \over 
                          \langle \tr[U_0({\bf 0},0) U_0^\dg({\bf R}_L,0)] \rangle }  -  \langle \oh \tr U_P\rangle \ , \non \\
\label{Qlat}
\eea
and where
\beq
            U_P = U_x(\vp,0) U_0(\vp+\ve_x,0)U_x^\dg(\vp,1)U^\dg_0(\vp,0)
\eeq 
is a plaquette operator, with
\beq
           \vp = y \ve_y + \ve_x \times \left\{ \begin{array}{cl}
                                      \oh R \mbox{~or~} \oh R -1 & ~~~~~R~\mbox{even} \cr
                                  \oh (R-1) & ~~~~~R~\mbox{odd} \end{array} \right.  \ ,
\eeq
and the expectation values are evaluated in Coulomb gauge.  The lattice operator in the numerator of \rf{Qlat} is illustrated in Fig.\ \ref{corr}.  Of course there is nothing special about the $x,y$-directions or
the $t=0$ time-slice, so in practice we average over observables which differ only by spacetime translations and 90${}^\circ$ spatial rotations (Coulomb gauge precludes changing the orientation in time).

\begin{figure*}[htb]
\subfigure[~]  
{   
 \label{r1}
 \includegraphics[scale=0.5]{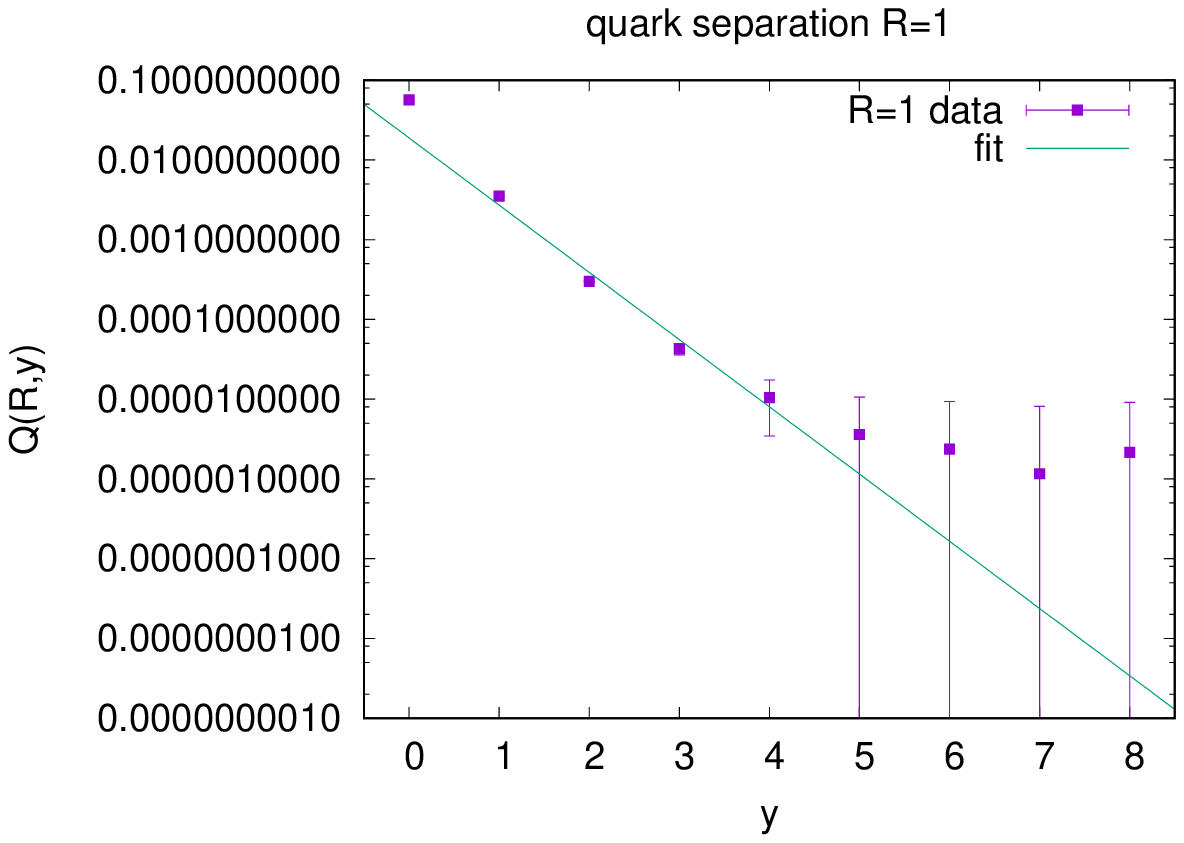}
}
\subfigure[~]  
{   
 \label{r2}
 \includegraphics[scale=0.5]{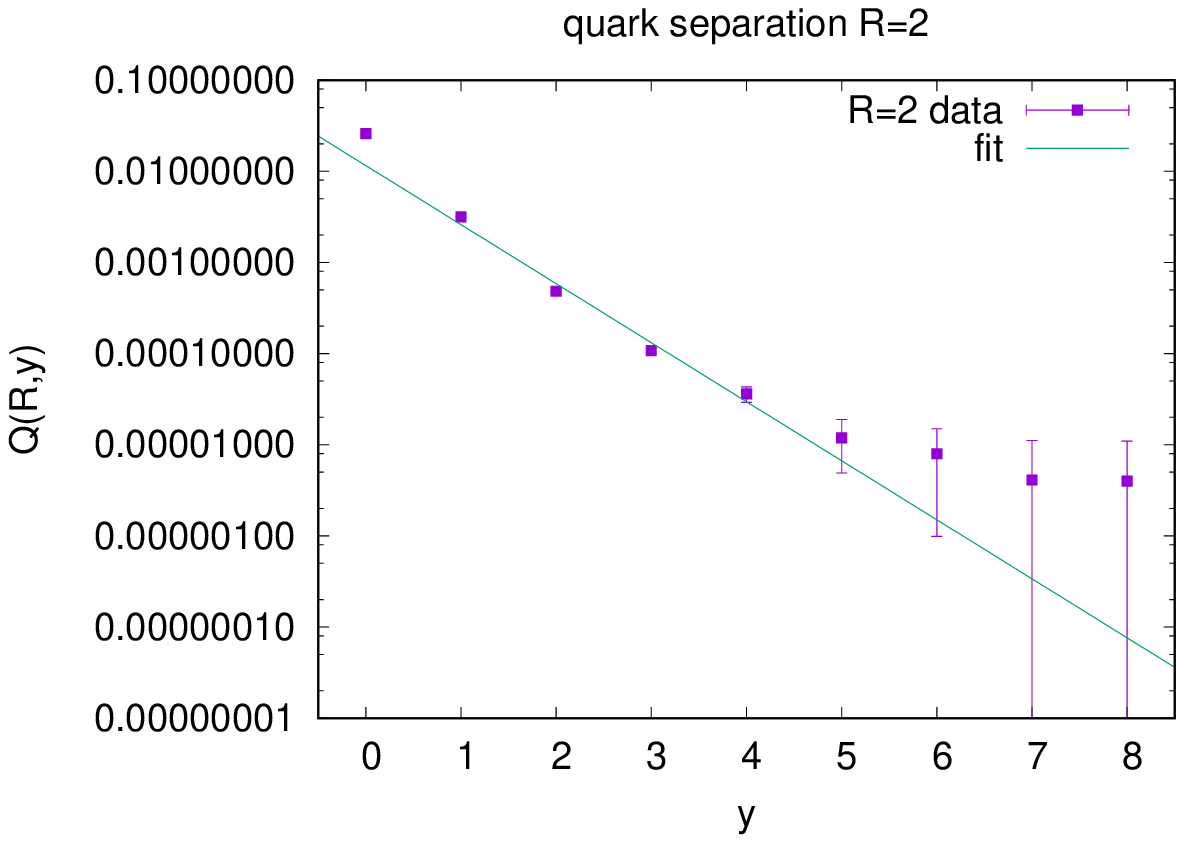}
}
\subfigure[~]  
{   
 \label{r3}
 \includegraphics[scale=0.5]{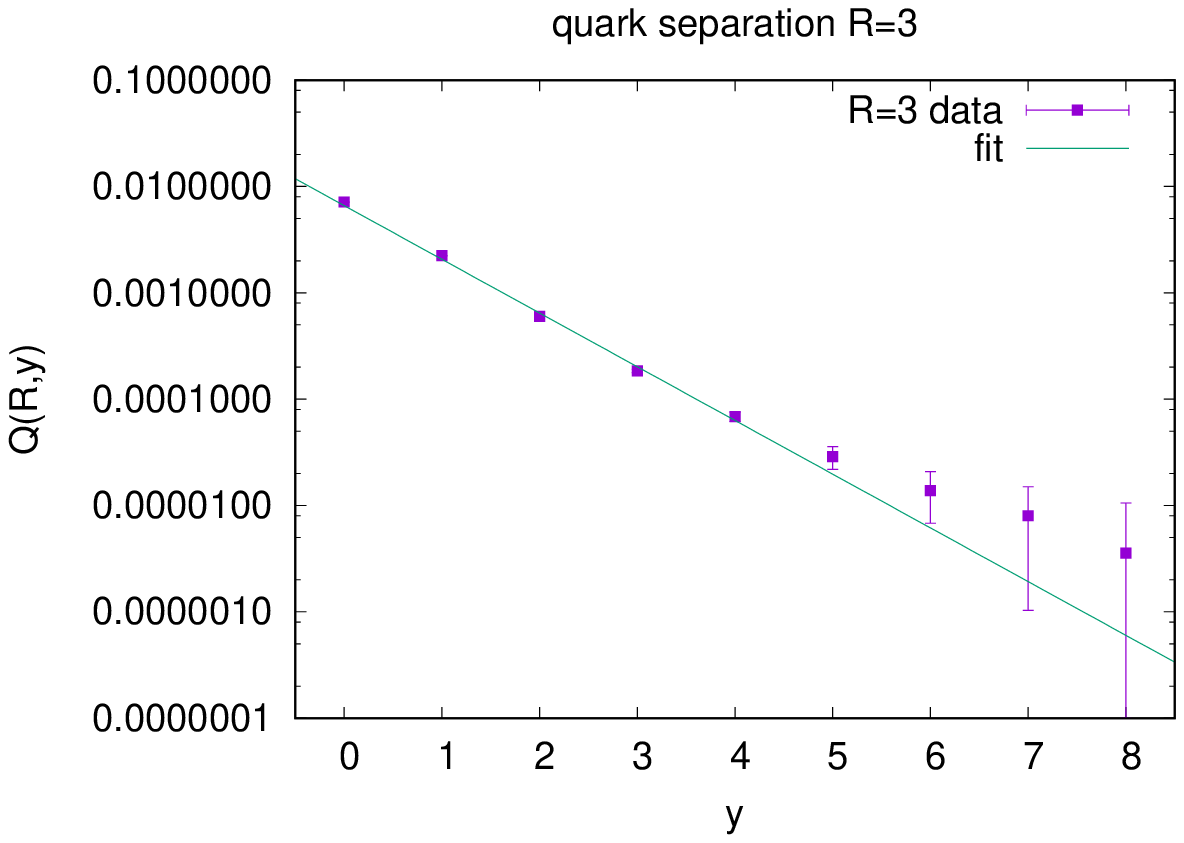}
}
\subfigure[~]  
{   
 \label{r4}
 \includegraphics[scale=0.5]{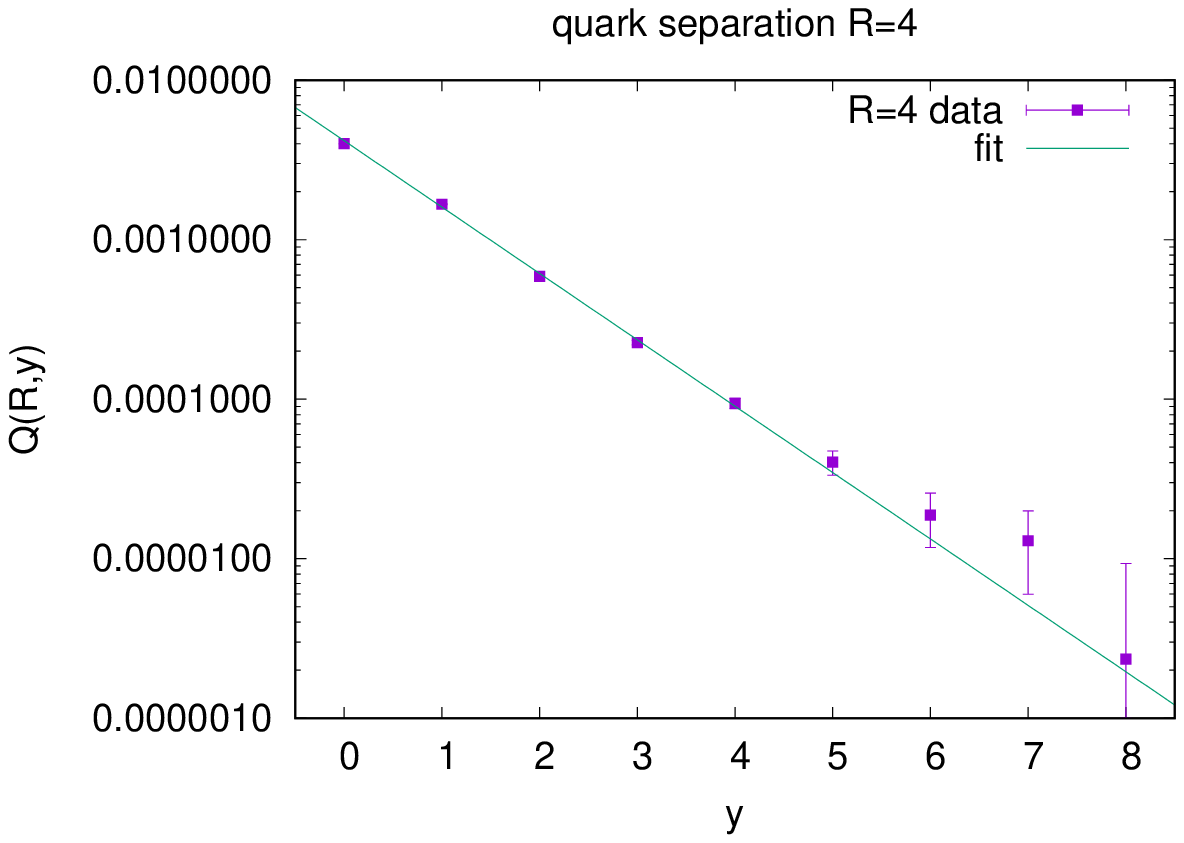}
}
\subfigure[~]  
{   
 \label{r5}
 \includegraphics[scale=0.5]{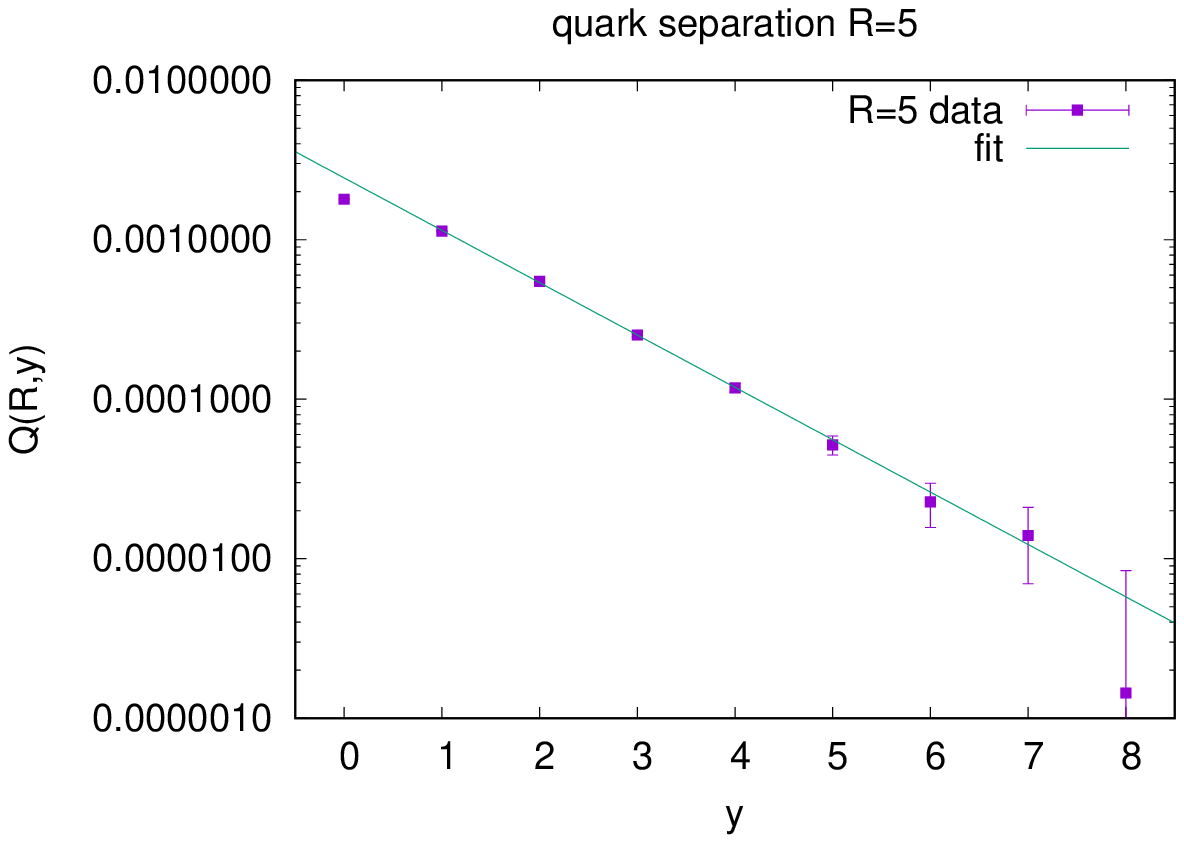}
}
\subfigure[~]  
{   
 \label{r6}
 \includegraphics[scale=0.5]{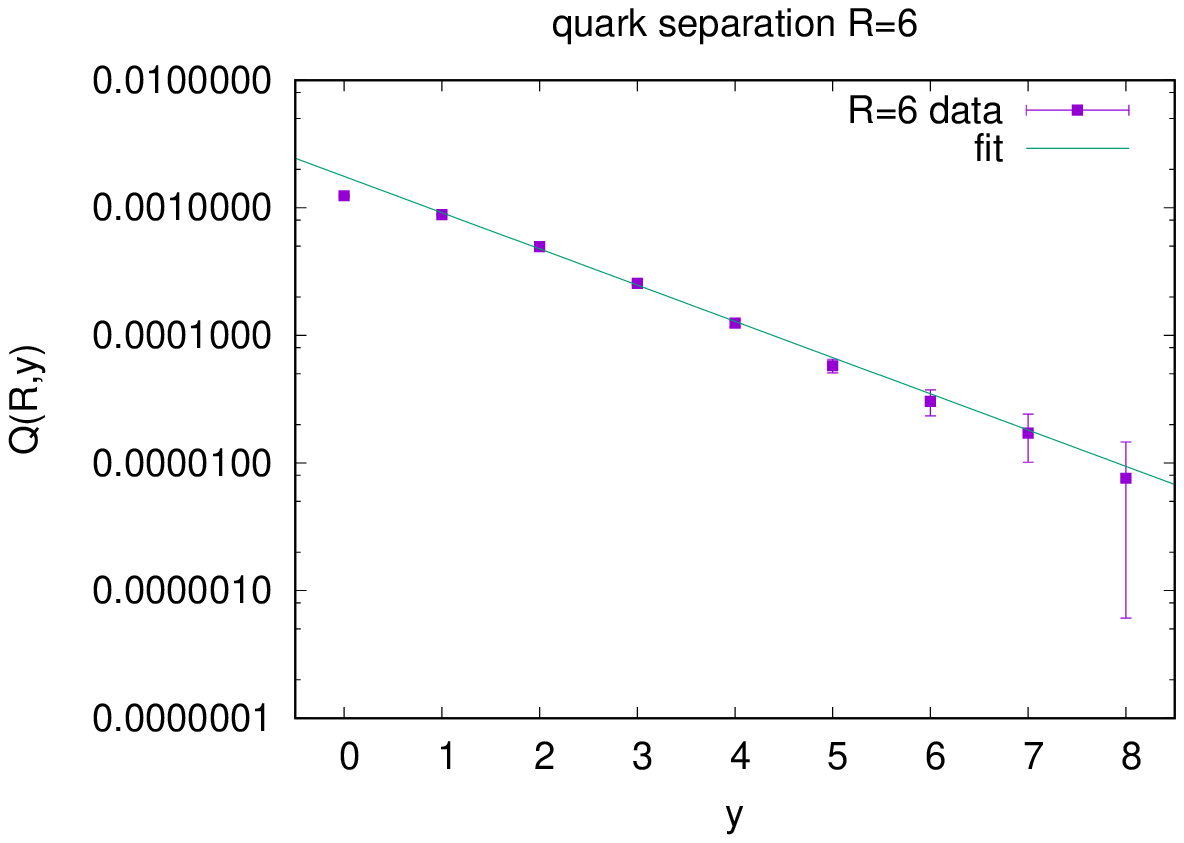}
}
\subfigure[~]  
{   
 \label{r7}
 \includegraphics[scale=0.5]{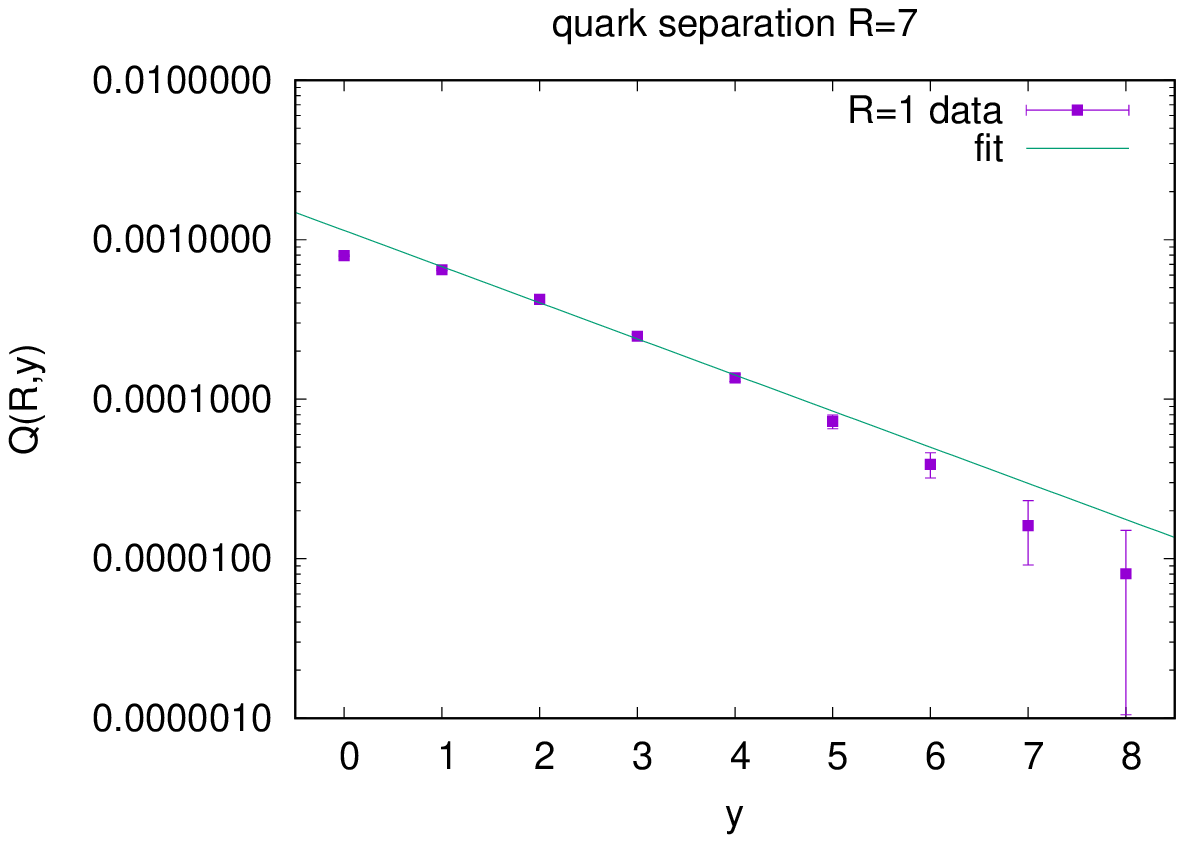}
}
\caption{The connected correlator $Q(R,y)$ of two timelike links and one plaquette, for fixed link separation $R$, vs.\ transverse
separation $y$ of the plaquette from the midpoint of the line of quark-antiquark separation. This is a measure of the
falloff of the color Coulomb energy density with transverse distance away from a quark-antiquark dipole in Coulomb gauge. The simulation is for SU(2) pure gauge theory at $\b=2.5$.  The lines
show a best fit to an exponential falloff $a\exp[-bR]$ in the range $1\le R \le 4$.}  
\label{Qdata}
\end{figure*}

\begin{figure*}[htb]
\subfigure[~]  
{   
 \label{anlge}
 \includegraphics[scale=0.17]{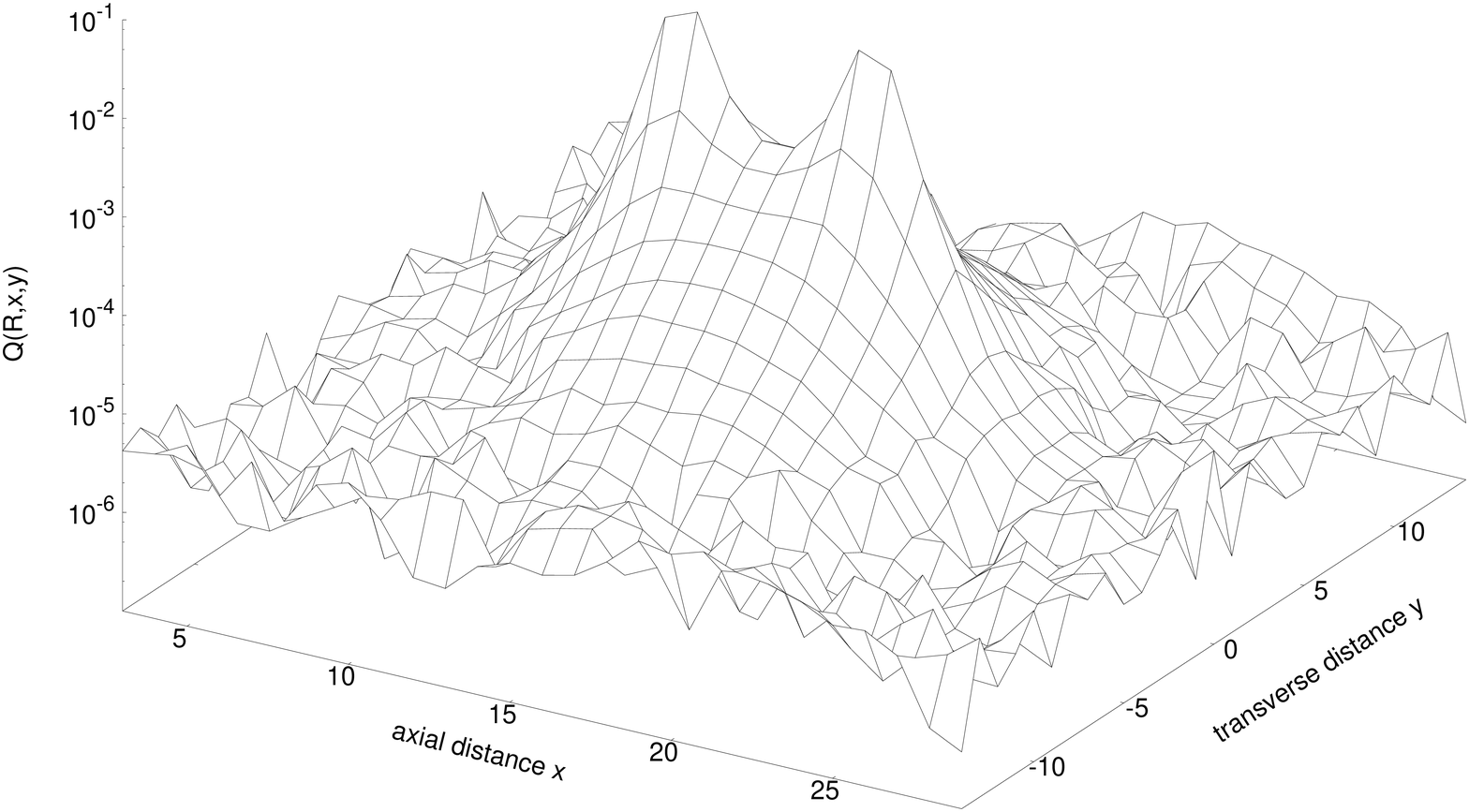}
}
\subfigure[~]
{   
 \label{edge}
 \includegraphics[scale=0.17]{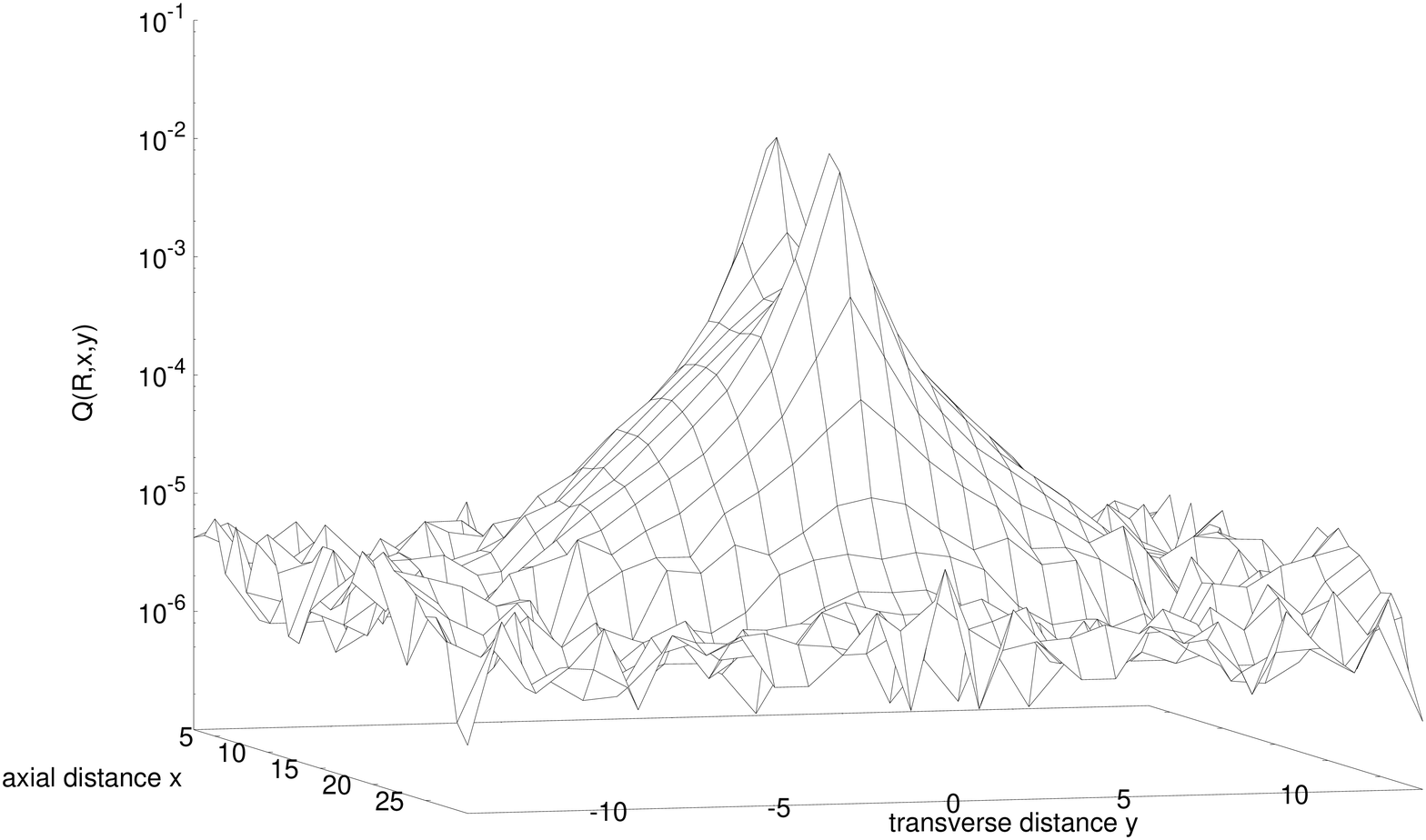}
}
\caption{Two views, from different perspectives, of the Coulomb flux tube at quark-antiquark separation $R=5$.  Note the
logarithmic scale on the $z$-axis.}
\label{3d}
\end{figure*}

\section{Results}

    Our results for quark-antiquark separations $1\le R\le 7$, and transverse separations $0\le y \le 8$, obtained from lattice Monte Carlo simulation of pure SU(2) gauge theory at $\b=2.5$ on a $24^4$ lattice volume, are shown in the logarithmic plots of Fig.\ \ref{Qdata}.  This data seems to rule out, fairly conclusively, any mild power-law falloff of the color electric energy density with
transverse distance $y$ from the midpoint.   The falloff with $y$ at fixed $R$ instead seems to be very nearly a pure exponential, at least until the error bars are comparable to the values of the data points.  The Coulomb electric field of the quark-antiquark dipole is therefore not long range, but rather is collimated along the axis of the quark-antiquark pair.  In other words, we appear to be seeing a Coulomb flux tube.  Comparison of Fig.\ \ref{r7} to a plot of the center plane action density in the asymptotic (or minimal energy) flux tube, shown in Fig.\ 14 of ref.\  \cite{Bali:1994de}, indicates that the Coulomb flux tube is substantially narrower than the minimal energy flux tube, with a width smaller by about a factor of 1.7.\footnote{Figure 14 in ref.\ \cite{Bali:1994de} was also taken at $\b=2.5$, for a quark-antiquark separation of 8 lattice spacings.}    This is an indication that the finite width of the Coulomb flux tube cannot simply be attributed to the finite
size of $t$ (in $L_t$) equal to the lattice spacing, in the lattice version \rf{Llogtime} of the correllator \rf{logtime}.  If it were the case that the width was infinite at $t \ra 0$ (i.e.\ power law falloff), and shrunk to the width of the minimal energy flux tube at $t \ra \infty$, then we would expect the width of the flux tube at finite lattice spacing to be {\it greater} than the width of the minimal energy flux tube, whereas the reverse is what we actually find.

    Profiles of the flux tube (or, more precisely, the component $\langle \oq  \tr E^2_x \rangle$) at a quark-antiquark separation of $R=5$ are shown in Fig.\ \ref{3d}.  In this case we are computing an observable $Q(R,x,y)$ defined by the right hand side of \rf{Qlat}, but with point $\vp$ defined by
\beq
             \vp = y \ve_y + x \ve_x \ .
\eeq
Note the logarithmic scale on the vertical axis of Fig.\ \ref{3d} .  On a linear scale, the values in the transverse direction are soon indistinguishable from zero.

\begin{figure}[htb]
\includegraphics[scale=0.7]{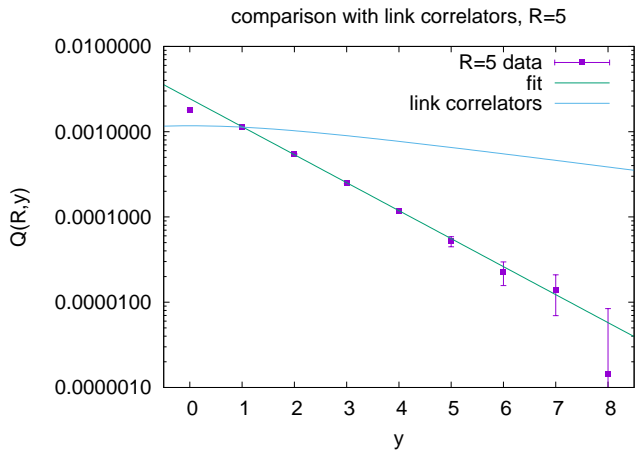}
\caption{The data for $Q(y,5)$, compared to the formula \rf{link_corr}, with a multiplicative constant $c_0$ chosen so that
the formula and data point agree at $y=1$.} 
\label{compare}
\end{figure}

   A natural question is whether the exponential falloff in the transverse direction is directly related to the exponential falloff of the timelike 
link correlator, from which we have extracted the color Coulomb potential.  For example, we might ask whether the connected correlator of the timelike links and timelike plaquette can be viewed as a four timelike link correlator, which factorizes into a product of two point functions.  Coulomb gauge brings spatial links as close as possible to the identity, so if we approximate 
\bea
         & &   U_x(\vp,0) U_0(\vp+\ve_x,0)U_x^\dg(\vp,1)U^\dg_0(\vp,0) \non \\
          & &  \qquad \qquad \approx U_0(\vp+\ve_x,0) U^\dg_0(\vp,0)  \ ,
\eea
then the numerator in \rf{Qlat} involves a four-point equal-time correlator of timelike links.  Let us suppose that the 
connected part approximately factorizes into a product of two-point functions
\bea
 & & \langle \tr[U_0({\bf 0},0) U_0^\dg({\bf R}_L,0)]  \tr[U_P(\vp,0)]\rangle_{conn} \non \\
  & & \qquad \sim \langle \tr[U_0({\bf 0},0) U^\dg_0(\vp+\ve_x,0)] \rangle \langle \tr [U_0^\dg({\bf R}_L,0) U_0(\vp,0)]\rangle \ . \non \\
\eea
In that case we would expect
\beq
    Q(y,R) \sim c_0  \exp\left[-\s_c\left(2\sqrt{y^2+\oq R^2} - R\right)\right]   \ ,
\label{link_corr}
\eeq
where $c_0$ is a constant.  In fact this is not even close to a fit of the data, as seen in Fig.\ \ref{compare} for $R=5$.  The exponential falloff of the Coulomb flux tube in the transverse direction is much faster than $\exp[-2\s_c y]$, and this two-point correlator description simply fails to give a reasonable account of the data. 

    Returning to the expression \rf{E2}, and the observed power law behavior of the ghost propagator \rf{ghost}, there is the question of how the Coulomb energy density could fail to also have a power law falloff.  The only possibility we see is that the ``typical'' vacuum configurations which account for the expectation value of the ghost propagator are not the dominant configurations in the expectation value of products of the $G^{ab}(\vx,\vy,A)$ operators, such as $ \nabla_x G^{ab}(\vx,0;A) \cdot \nabla_x G^{ab}(\vx,R;A)$.  The expectation value of the product must be very sensitive to exceptional field configurations, possibly ones in which the lowest eigenvalue of the Faddeev-Popov operator is unusually small, which do not greatly affect the expectation value of the operator $G^{ab}(\vx,\vy,A)$
by itself.  Presumably these exceptional configurations, for reasons that are not clear to us, must be responsible for the rather precise cancellations among the different terms in \rf{E2} that are required for an exponential falloff.

\bigskip

\section{Conclusions}

   Confinement in Coulomb gauge seems to be more subtle than simply a linear potential from dressed one-gluon exchange, i.e.\ 
$\langle A_4 A_4 \rangle$.  While this two point function is no doubt part of the story, it is not the whole story, since the two point function, although linearly rising, is not by itself an explanation for the formation of a Coulomb electric flux tube.  One may speculate that the Coulomb string tension derives from the same underlying mechanism (center vortices come to mind) that accounts for the asymptotic string tension.  In particular, the confining two point function is thought be due to a non-perturbative enhancement in the density of near-zero eigenvalues of the Faddeev-Popov operator, associated with the proximity of typical vacuum configurations to the Gribov horizon.   It has been shown via lattice Monte Carlo simulations that removal of center vortices from thermalized lattice configurations sends this eigenvalue density back to the perturbative form \cite{Greensite:2004ur}, and the corresponding Coulomb string tension (along with the asymptotic string tension) vanishes upon vortex removal  \cite{Greensite:2003xf} (for recent developments in the vortex picture, see \cite{Trewartha:2015nna,*Kamleh:2017lij}).  If center vortices or some other topological objects are responsible for the linearly rising Coulomb potential, it is probably necessary to also appeal to a topological mechanism in order to understand the formation of a Coulomb flux tube.   

\bigskip

\acknowledgements{This research is supported by the U.S.\ Department of Energy under Grant No.\ DE-SC0013682.}

\bibliography{chain}
\end{document}